# IterMiUnet: A lightweight architecture for automatic blood vessel segmentation

Ashish Kumar[1,2], R. K. Agrawal[1], Leve Joseph[2]

1 Department of Computer Science and Engineering, Indian Institute of Technology Roorkee, India

2 School of Computer and Systems Sciences, Jawaharlal Nehru University, New Delhi, India

3 All India Institute of Medical Sciences, New Delhi, India

## Abstract

The automatic segmentation of blood vessels in fundus images can help analyze the condition of retinal vasculature, which is crucial for identifying various systemic diseases like hypertension, diabetes, etc. Despite the success of Deep Learning-based models in this segmentation task, most of them are heavily parametrized and thus have limited use in practical applications. This paper proposes IterMiUnet, a new lightweight convolution-based segmentation model that requires significantly fewer parameters and yet delivers performance similar to existing models. The model makes use of the excellent segmentation capabilities of Iternet [1] architecture but overcomes its heavily parametrized nature by incorporating the encoder-decoder structure of MiUnet [2] model within it. Thus, the new model reduces parameters without any compromise with the network's depth, which is necessary to learn abstract hierarchical concepts in deep models. This lightweight segmentation model speeds up training and inference time and is potentially helpful in the medical domain where data is scarce and, therefore, heavily parametrized models tend to overfit. The proposed model was evaluated on three publicly available datasets: DRIVE, STARE, and CHASE-DB1. Further cross-training and inter-rater variability evaluations have also been performed. The proposed model has a lot of potential to be utilized as a tool for the early diagnosis of many diseases.

## Keywords

## Declarations

**Funding:** Not Available

**Conflicts of interest/Competing interests:** No conflict of interest

**Availability of data and material:** All datasets are publicly available

DRIVE- https://drive.grand-challenge.org/   STARE-https://cecas.clemson.edu/~ahoover/stare/probing/index.html

CHASE DB1- https://blogs.kingston.ac.uk/retinal/chasedb1/

## 1. Introduction

Fundus imaging of the retina is a non-invasive and cost-effective way of diagnosing various ophthalmological and systemic diseases like high blood pressure, hypertension, diabetes, etc. [3]. The early manifestations of various cardiovascular diseases appear as changes in the structure of blood vessels which can easily be diagnosed by analyzing the fundus images [4]. Segmenting the blood vessel structures is considered the fundamental step toward quantitative analysis of these fundus images. The presence of pathologies such as microaneurysms and hemorrhage, low contrast, and variable vessel width in the fundus images makes this segmentation a non-trivial task. Currently, the segmentation process is being done manually by human annotators. However, manual segmentation is a long and laborious process [5]. In addition, it is heavily dependent on the subjective skills of annotators. The issues involved in the imaging procedure, such as non-uniform background illumination, poor contrast between vessels and the background, and variations in vessel shape and width, limit the degree of overlap between the segmentation performed by various human annotators. Automatic vessel segmentation is a natural replacement for the costly manual segmentation process. The ability to automatically segment blood vessel structures can aid in analysis, disease screening, and rapid detection of underlying diseases such as glaucoma, age-related macular degeneration, and diabetic retinopathy. Incorporating automatic segmentation methods into computer-aided diagnostic (CAD) systems can facilitate the development of a fully automated diagnosis system. These CAD systems can assist ophthalmologists and accommodate the rising number of patients in the health system. As a result, developing efficient algorithms capable of autonomously segmenting the vascular structure is an important research area.

Unet [6] has already been used to conduct biomedical image segmentation with success. It is made up of two symmetrical modules, an encoder, and a decoder, which is aligned in a U-shape. The encoder portion, which consists of stacks of convolutions and pooling layers, extracts the features from the image. The decoder section consisting of upsampling operators reconstructs the segmentation map from the feature maps extracted by the encoder section. Skip connections are present in Unet that merge low-level feature maps from the encoder layer to the mirrored high-level feature maps of the decoder layer to enable precise localization in the segmented image. However, the complexities of vessel structures present in fundus images pose a significant challenge for the plain vanilla Unet. It fails to segment accurately, particularly in the regions marked by thin vessels. Several other variants [1, 2, 7, 8] of Unet have been proposed to solve this problem with limited success. Even so, the use of Unet and its existing variants is limited because they are heavily parametrized, making them inefficient for practical applications.

We present IterMiUnet, a fully convolutional model, in this paper. The model is a variant of Unet but overcomes the problems of Unet based segmentation. Specifically, we have worked to get around the heavy parameterization problem of Unet and its variants which is a huge bottleneck in their usage. The proposed model is inspired by two existing models, namely Iternet and MiUnet. We replaced Iternet's encoder-decoder structure with MiUnet's lightweight blocks. This incorporation of MiUnet blocks into the Iternet architecture significantly decreases parameters while preserving the original depth. Since the depth of the model remains unchanged, we can get around the regions of thin vessels as effectively as in the Iternet architecture but with the light-weighted efficiency of the MiUnet model. Our proposed lightweight model with roughly about 0.15 M parameters is a far cry from the existing data-hungry, overly parametrized models that are notorious for overfitting. It results in a significant reduction in training and inference time and thus can quickly be adopted for practical use cases. We have included Unet and its various variants in our study to evaluate its performance. Even with such a low parametrization, the performance of our proposed method matches the best of the variants of the Unet model. The remaining sections of this work are organized in the following way: A short literature review of related papers is presented in Section 2. The proposed model is discussed in Section 3. Section 4 presents experimental results to evaluate the proposed model against various Unet-based models and existing segmentation methods across multiple datasets, followed by a discussion. Section 5 describes the conclusions drawn from this research.

## 2. Related Work

In the vessel segmentation problem, the end goal is to separate the blood vessels from the retinal images. This is done by locating and identifying retinal vessel structures. The methods used to achieve this objective can be classified into two categories: unsupervised and supervised. Table 1 summarises all the approaches reviewed in this paper, followed by a brief overview of these methods in the following subsections.

Table 1 Summary of related papers. (Following abbreviations are used: SE-Sensitivity, SP-Specificity, AC-Accuracy)

| Authors | Publication Year | Methodology | Dataset | Evaluation Metrics | Category |
|---|---|---|---|---|---|
| Chanwimaluang et al.[9] | 2003 | Local Entropy Thresholding | STARE | Visual comparison with groundtruth | Unsupervised Methods |
| Martinez Perez et al. [10] | 2007 | Multi-Pass Region Growing | DRIVE, STARE | SE, SP, AC | |
| Lam et al. [11] | 2010 | Multi Concavity approach | DRIVE, STARE | AC, AUC | |
| Fraz et al. [12] | 2012 | Morphological approach | DRIVE, STARE, MESSIDOR | SE, SP, AC | |
| Roychowdhury et al. [13] | 2015 | Adaptive thresholding | DRIVE, STARE, CHASE_DB1 | SE, SP, AC, AUC | |
| Azzopardi et al. [14] | 2015 | COSFIRE filter | DRIVE, STARE, CHASE_DB1 | SE, SP, AC, AUC | |
| Fan et al. [15] | 2019 | Hierarchical Image Matting | DRIVE, STARE, CHASE_DB1 | SE, SP, AC, AUC | |
| Tian et al. [16] | 2021 | Frangi filter | DRIVE, STARE | SE, SP, AC | |
| Ricci et al. [17] | 2007 | Support Vector Machine | DRIVE, STARE | AC, AUC | Traditional Machine Learning Methods |
| Marin et al. [18] | 2011 | Neural Network classifier | DRIVE, STARE | SE, SP, AC, AUC | |
| Fraz et al. [19] | 2012 | Ensemble classifiers | DRIVE, STARE, CHASE_DB1 | SE, SP, AC, AUC | |
| Roychowdhury et al. [20] | 2015 | Gaussian Mixture Model classifier | DRIVE, STARE, CHASE_DB1 | SE, SP, AC, AUC | |
| Li et al. [21] | 2016 | Deep Learning model | DRIVE, STARE, CHASE_DB1 | SE, SP, AC, AUC | Deep Learning methods |
| Liskowski et al [22] | 2016 | Deep Learning model | DRIVE, STARE | AC, AUC | |
| Oliveira et al. [23] | 2018 | Deep Learning model | DRIVE, STARE, CHASE_DB1 | SE, SP, AC, AUC | |
| Yan et al. [24] | 2018 | Unet with joint loss | DRIVE, STARE, CHASE_DB1 | SE, SP, AC, AUC | |
| Alom et al. [7] | 2018 | Recurrent Unet | DRIVE, STARE, CHASE_DB1 | SE, SP, AC, AUC | |
| Jin et al. [8] | 2019 | Deformable Unet | DRIVE, STARE, CHASE_DB1 | SE, SP, AC, AUC | |
| Hu et al. [2] | 2019 | MiUnet | DRIVE, STARE, CHASE_DB1 | SE, SP, AC, AUC | |
| Li et al. [1] | 2020 | Iternet | DRIVE, STARE, CHASE_DB1 | SE, SP, AC, AUC | |
| Mou et al. [25] | 2021 | Unet with Self Attention | DRIVE, STARE | SE, SP, AC, AUC | |
| **Proposed Model** | **2022** | **IterMiUnet** | **DRIVE, STARE, CHASE_DB1** | **SE, SP, AC, AUC, F1** | |

## 2.1 Unsupervised methods

Unsupervised methods, in general, do not rely on manual annotation data. They use pre-defined rules such as vessel tracking, matched filtering response, and morphological processing to extract vascular structures and achieve segmentation. Chanwimaluang et al. [9] suggested a four-step approach that included matched filtering, entropy-based thresholding, length filtering, and detection of vascular intersections. However, the method involves longer computational steps. Martinez Perez et al. [10] employed a multiscale feature extraction followed by a multi-pass region growing methodology to segment retinal blood vessels autonomously. Lam et al. [11] devised a multi-concavity approach to segment both healthy and pathological images. Fraz et al. [12] employed a Gaussian filter first-order derivative to identify vessel centerlines in fundus images and paired it with morphological bit plane slicing to extract vessels. Roychowdhury et al. [13] introduced a new iterative unsupervised approach based on an adaptive thresholding method using morphologically enhanced vessel images. Azzopardi et al. [14] developed a modified B-COSFIRE filter that can detect retinal vessels using the difference-of-Gaussian (DoG) filter and mean operation shifting. Fan et al. [15] used a hierarchical image matting methodology to extract blood vessels from a fundus image. Tian et al. [16] employed a modified Frangi filter algorithm to improve the visibility of blood vessel structures and a mathematical morphology method to remove noise interference around the thin vessel structures. The fixed set of segmentation rules learned by various unsupervised methods mentioned above fails to capture the diverse vessel complexities and does not generalize well across multiple datasets. Since unsupervised methods' performance is lower than supervised methods, many researchers are exploring and adopting supervised methods.

**2.2 Supervised methods**

These methods require knowledge of the groundtruth of training images in advance. They work by learning a mapping between the input image and the groundtruth, which is subsequently used to classify every pixel in the image as a vessel or background pixel. Traditional supervised learning methods relied on handcrafted features to characterize a pixel that is fed to a classifier model that learns a mapping function. Ricci et al. [17] combined gray levels of pixels and line detectors to obtain a feature vector which is passed to a support vector machine for pixel classification, thereby creating a vessel map. Marin et al. [18] suggested a 7D feature vector based on grey level and moment invariants-based features for blood vessel extraction from preprocessed retinal images. The feature vector is subsequently fed into a neural network for pixel classification. Fraz et al. [19] used a feature vector constructed by combining gradient vector field orientation analysis, morphological transformation, line strength metrics, and Gabor filter responses. This feature vector is passed to an ensemble of bagged and boosted decision trees. A Gaussian mixture model is used by Roychowdhury et al. [20], which gets its input in the form of an eight-dimensional feature vector based on each pixel's immediate surroundings and gradient images of first and second order. Thick vessels and thin vessel regions were segmented separately in this manner. Each of the above methods employed handcrafted feature descriptors, and the quality of extracted features directly impacts classification accuracy. As a result, these methods are complex and lack robustness, which invariably results in poor generalisation. The segmentation performance is dependent on the quality of features which inevitably reaches a bottleneck due to subjectivity of humans involved in feature selection.

Traditional machine learning approaches rely on feature selection after careful extraction of features. However, feature extraction is a time-consuming and labor-intensive process that necessitates greater prior information. Deep learning algorithms have brought a paradigm shift as they can automatically learn features using a deep hierarchical feature extraction approach [26] instead of relying on handcrafted features. Although the basic technical idea behind deep learning has been known for decades [27], it has only recently been applied to various fields, including retinal vessel segmentation. Several deep learning models based on Convolutional Neural Networks (CNNs) have been suggested for the segmentation of retinal vessels. Li et al. [21] used a deep neural network to represent the image to segmentation map transformation as a cross-modality data transformation. Liskowski et al. [22] tested several variants of deep learning networks on a large dataset generated via data augmentation, which also includes a scheme to highlight context information called structured prediction. However, this model's data preprocessing and enhancement methods were overly complex, and the algorithm did not perform well in the segmentation of thin vascular branches. Oliveira et al. [23] used wavelets to increase input images' channels and multiscale CNN to segment retinal vessels. In the research work [24], a new loss function is proposed for the convolutional networks, emphasizing the segmentation accuracy of thin vessels that are otherwise missed by the standard pixel-level loss function. Long et al. [28] pioneered fully convolutional networks, replacing densely connected layers with convolutions. This network can take images of any size as input and output a segmentation map with the exact dimensions. The concept of combining feature maps at various levels of abstraction was also explored. Several multiscale architectures for semantic segmentation have been suggested [29, 30]. Following this trend, Unet [6] was introduced for biomedical segmentation. Unet and its variants uses an encoder-decoder structure for segmentation. In [7], Recurrent Unet was introduced for retinal vessel segmentation. The model consists of recurrent convolution blocks characterized by time unfolding convolutions, thus ensuring a deeper and more precise feature representation. Jin et al. [8] introduced deformable convolution blocks which are characterized by a dynamic, receptive field. MiUnet model, introduced in [2], is a simplified Unet architecture but generates a highly inaccurate vessel segmentation map across various datasets. Iternet, as proposed in [1] determines a vessel segmentation map using a similar encoder-decoder structure to Unet and then refines this imprecise map using iterative modules. Mou et al. [25] proposed a model with increased hierarchical representation capture ability that makes use of self-attention mechanism in the U-shape encoder-decoder.

Extensive research has been done into retinal vessel segmentation, yet the need for an accurate vessel segmentation map is far from over. The various variants of Unet model forms the state-of-the-art for vessel segmentation. These models make use of innovative modifications to the encoder-decoder structure of the original Unet model to get around the complex vessel structures of fundus images. These modifications have made the Unet variants excessively complicated and overly parametrized. Although deep learning methods invariably perform better than traditional machine learning approaches, the segmentation maps generated still have visible defects. Thus, proposing a simple and effective model for vessel segmentation is still an open problem. This paper addresses this problem by proposing a simple, lightweight segmentation model for automatic retinal blood vessel segmentation.

## 3. IterMiUnet: A lightweight architecture

Unet is a segmentation model based on fully convolutional networks. It consists of two modules: an encoder and a decoder. The encoder module forms the contracting path of the architecture, where stacks of convolution (conv) and max pooling layers are successively applied to the input image to obtain low-resolution, highly abstract features. This successive downsampling of the image in the contracting path helps capture the context of the image but results in the loss of location information regarding various image objects. However, to reconstruct the segmentation map, information on the location of each object in the image is also necessary. This is where the decoder module comes into play. The decoder module forms the expanding path of the architecture, where stacks of upsampling operators and convolution layers are successively applied to the low-resolution feature maps. This successive upsampling of the feature maps in the expanding path recovers location information. To ensure precise localization of objects, every level in the encoder module is connected to a mirrored level in the decoder module with the help of skip connections. These skip connections merge low-level information from shallow layers to high-level information present in deep layers. Fig. 1 depicts a typical Unet architecture where a segmentation map for a grayscale image of size $48 \times 48$ is obtained. The number of filters at each successive level is also displayed. The upsampling operators in the expanding path are denoted by transposed convolution (Trans Conv) in Fig. 1.

Unet has already proven its worth for segmentation tasks. However, its usefulness in retinal vessel segmentation is severely limited for the following reasons: First, the increase in depth increases the number of learnable parameters but does not necessarily increase new discriminative features. Secondly, the intricate complexities of the blood vessels pose a significant challenge for Unet, due to which it produces a segmentation map with noticeable errors and low sensitivity. Several variants have been tried to deal with this problem. However, heavy parameterization remains a significant bottleneck that limits the use of any Unet model for real-world applications.

We propose IterMiUnet, a new model that can effectively segment the blood vessels with significantly fewer learnable parameters. We achieved this by combining two existing models, namely MiUnet and Iternet. MiUnet and Iternet have been described separately in the below subsections to understand the proposed new model better.

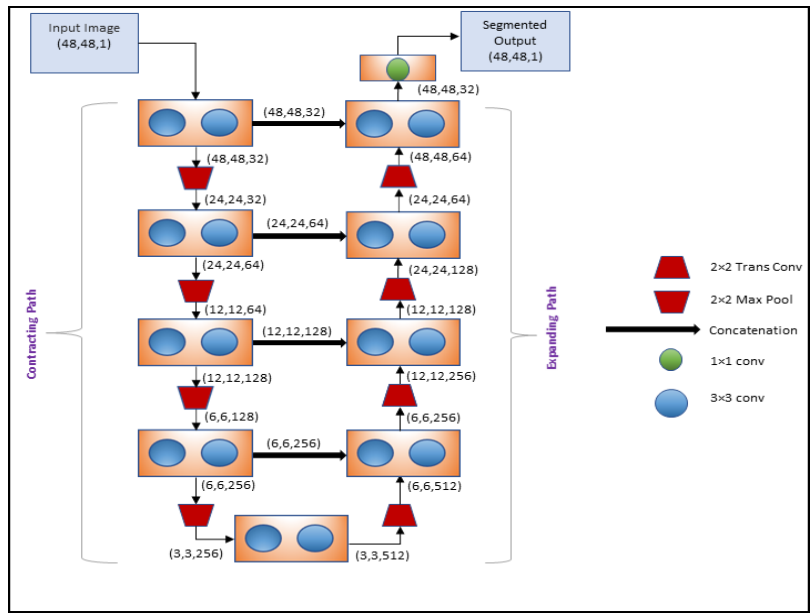

**Fig. 1** Architecture of Unet model.

### 3.1 MiUnet

The usual pipeline for an encoder-decoder based architecture consists of an increasing number of filters at each subsequent level in the contracting path and vice versa in the expanding path. MiUnet reverses this filter configuration while maintaining the same depth as a Unet model. Thus, in a MiUnet architecture, the filter count in the expanding path decreases or stays the same, and vice versa occurs in the expanding path. This filter configuration in MiUnet allows us to have a deeper network without a significant increase in parameters. Fig. 2 depicts a typical MiUnet architecture, showing the number of filters at various levels and a final segmented output for a 48 × 48 grayscale image.

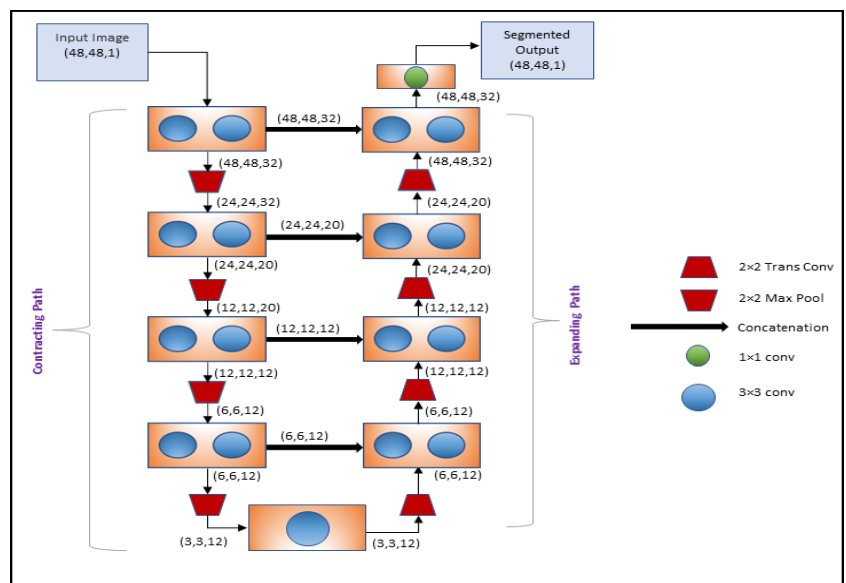

**Fig. 2** Architecture of MiUnet model

### 3.2 Iternet

Iternet is another variant of Unet which tries not to alter the feature extraction capability of Unet. It consists of a base module followed by an iterative module. The base module consists of the usual Unet architecture without any modification. The base module does most of the feature extraction and segmented output reconstruction, but the output at this stage is still an imprecise segmentation map. Therefore, an iterative module consisting of small mini Unet is introduced to refine the segmented map of the base module. The encoder-decoder structure inside the iterative module is exactly like Unet but with much lesser depth, thus being called the mini Unet. The iterative module acts like a refinery module whose primary function is to refine the feature map which invariably has noticeable errors. Each time a refinery module is run, it inputs the feature map from the previous module's second last layer. This enables us to have more information because the input is a multi-channel feature map instead of a single-channel vessel probability map. Additional skip connections exist connecting the base module to the mini Unet of each refinery module. This skip connection allows each refinery module to access the low-level features from the base module, which are remarkably close to the raw input image. The iterative working of the refinery module results in the refinery module receiving different inputs and producing varying segmentation maps after each iteration. As shown in Fig 3. the iterative module runs a refinery module at each iteration to obtain a segmentation map. Output $N$ will give us our final segmentation map after running the iterative module $N-1$ times. Iterative mini Unet execution in this segmentation process exposes the refinery modules to new failure patterns, resulting in a cleaner segmentation map.

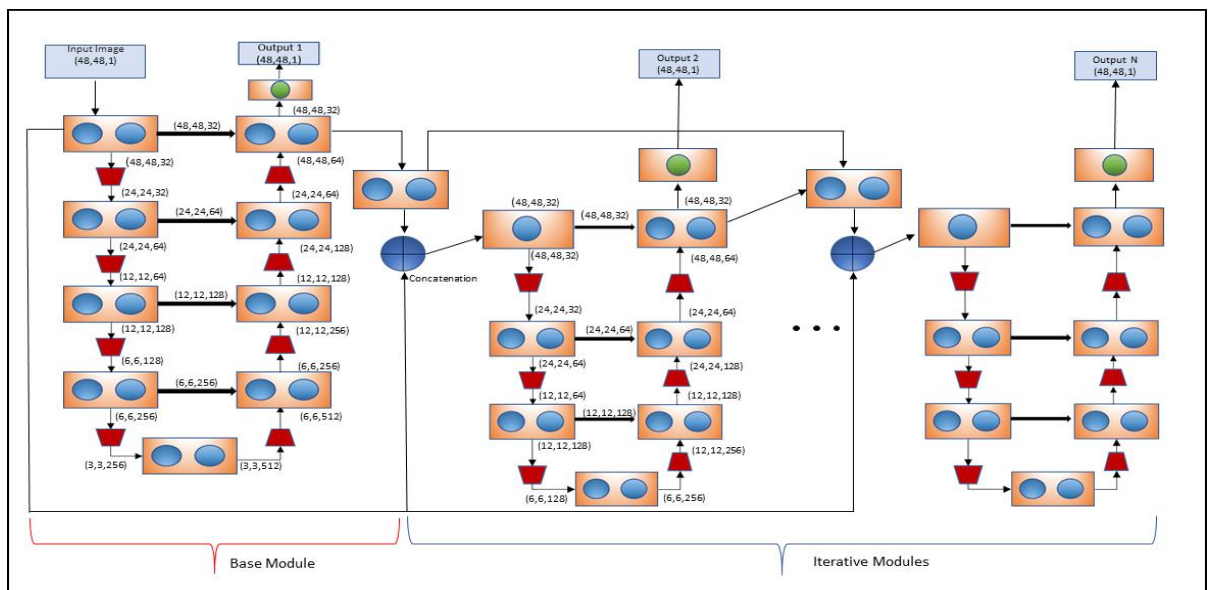

**Fig. 3** Architecture of Iternet model

### 3.3 Proposed IterMiUnet model

The Iternet, as explained in the above section, illustrates the concept of using an iterative model, which is a straightforward yet effective way to improve the vessel map without altering the structure of Unet. It generates a much cleaner segmentation map of the retinal images and overcomes many of the observed errors that appear in the segmentation map of a simple Unet. Although Iternet shows superior performance to Unet, it does so at the expense of a massive surge in the number of learnable parameters and a significant increase in training time. The use of mini Unet in an iterative module increases the model's parameters, leading to overfitting. On the other hand, MiUnet offers a lightly parametrized model. MiUnet achieves this by reducing the number of parameters in the contracting path and vice versa in the expanding path. Although this decreases learnable parameters, it also limits the number of abstract high-level concepts that a model can learn during the feature extraction process. This is evident from MiUnet's segmentation map, which shows a significant decline in segmentation performance. To summarize, Unet and its variants where the filters increase with depth in the contracting feature extraction path tend to overfit and are characterized by higher training time. The lightweight Unet-based models where filters decrease with the depth in the contracting feature extraction path have fewer learnable parameters and lower training time but are characterized by poorer segmentation performance.

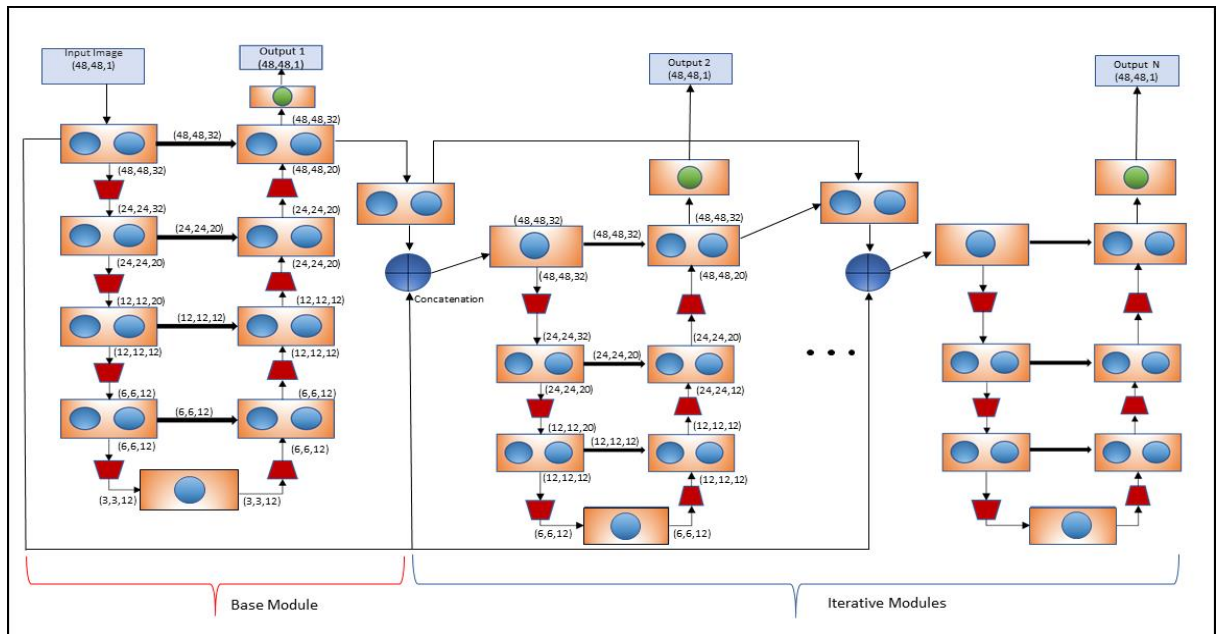

**Fig. 4** Architecture of IterMiUnet model

This paper presents IterMiUnet, a new segmentation model based on U-shaped architecture that takes care of the above discussed problems of Unet and its variants. This lightweight model is formed by combining Iternet with MiUnet. Fig. 4 shows the architecture of IterMiUnet. The number of filters at each level are displayed in the figure. It consists of two main working components: the base and iterative modules.

The base module of IterMiunet has the same functionality as the base module of Iternet. It does most of the feature extraction and segmented output reconstruction. However, we use the lightweight MiUnet as our architecture for the base module instead of Unet. It precisely means that we have adopted a filter configuration where the number of filters reduces or stays the same with the increase in depth during the contracting path and vice versa occurs in the expanding path. This obviously reduces the number of parameters but can also limit the discovery of abstract high-level features required to reconstruct a precise segmentation map. In fact, this has been a significant drawback that restricted the success of MiUnet in the segmentation of blood vessels. IterMiUnet, on the other hand, is equipped with an iterative module, which enables it to overcome the drawbacks of the MiUnet model despite using the same filter configuration.

The functionality of the iterative module of IterMiunet is identical to that of Iternet. It acts as a refinery module and refines the imprecise segmentation map received from the previous module at each iteration. We replaced the encoder-decoder structure inside the refinery module with mini MiUnet i.e., it is architecturally similar to MiUnet but with much lesser depth. Thus, contrary to Iternet, which employs the Unet architecture, we use the lightweight MiUnet architecture throughout. The placement of skip connections, the access to low-level features from the base module, and other functionalities of the iterative module remain unaltered. Despite gaining access to only a limited number of features extracted from the base module, the successive use of refinery modules at each iteration improves the segmentation map and makes it closer to groundtruth. Thus, it can be said that the iterative module of IterMiUnet is better able to capture the defects prevalent in the base module and produce a clean segmentation map.

The unreasonable effectiveness of the IterMiunet model can be gauged by the fact that the model can produce a vessel map with good segmentation performance despite the limited number of parameters, which restricts the number of high-level features that can be learned. Thus, integrating these two existing models produces a new lightweight model that can effectively segment the vessel structures in retinal images.

## 4. Results and Discussions

To check the performance of the proposed IterMiUnet model in comparison to Unet and its variants, we performed experiments on three publicly available datasets. This section contains all the details of the experiments, followed by a discussion.

## 4.1. Datasets

The DRIVE [31] (Digital Retinal Images for Vessel Extraction) is the most popular publicly available dataset for retinal images. The dataset was obtained as a consequence of a screening program that was carried out on 400 people in the Netherlands. The images have a resolution of 565×584 pixels. The images are separated into two sets, each with 20 images: training and test.

The STARE [32] consists of 20 fundus images, each consisting of 700×605 pixels in resolution. This dataset, unlike DRIVE, does not have separate train and test sets. The first half consisted of images of healthy subjects, while the other half had several abnormalities and pathologies which overlapped with the blood vessel structures. The segmentation becomes more challenging due to severe clinical complications in the STARE images, thus providing a more realistic performance evaluation condition.

The CHASE-DB1 [33] dataset contains images of the retina obtained from screening programs in England conducted on multi-ethnic children. The images are 999×960 pixels in size. There are 28 images with no specific distinction between the train and the test set. The images have low blood vessel contrast, non-uniform illumination in the background, and a bright strip of light running down wide arterioles, which are called the Central Vessel Reflex. Compared to DRIVE and STARE, this dataset is considered more challenging to segment.

The first observer's manual annotations are considered as the groundtruth in all the three datasets. For STARE and CHASE-DB1 there are no FOV masks available, and they have been generated according to the method given in [8]. STARE and CHASE-DB1 do not have different train and test sets. Therefore, for STARE, we employed the leave-one-out technique, while in CHASE-DB1, the first fourteen images made up the training set and the remaining fourteen made up the testing set.

## 4.2 Preprocessing and Patches Extraction

Deep neural networks can learn from raw fundus images, but they perform better when applied with good preprocessing strategies. We converted the raw RGB images into grayscale images to get better vessel-background contrast. To further improve the foreground and background contrast, the entire dataset was subjected to Normalization and Contrast Limited Adaptive Histogram Equalization [34]. Gamma correction was subsequently used to improve image quality. We have trained models on small patches to overcome the data scarcity problem and reduce overfitting. Each patch has a fixed dimension and is obtained by selecting a random centre within the entire image. We also select the patches that lie completely or partially out of the FOV, which helps the network to distinguish blood vessels from the FOV borders.

The optimum size for the patch needs to be carefully chosen because a smaller patch size tends to learn the local details and a smaller context central to that patch alone instead of the global details of the image. In contrast, a bigger patch size will increase the calculation complexity. We set the patch size to 48×48 as used in the research work [5]. A training set, a validation set, and a test set are generated from the dataset. 10,000 patches are randomly sampled from each image during training, with 1,000 used for validation. The total training patches in DRIVE, STARE, and CHASE-DB1 are 180000, 171000, and 126000, respectively, and the total validation patches are 20000, 19000, and 14000, respectively. During testing, overlapping patches are extracted, and the vessel's probability is calculated by averaging the predictions of multiple overlapping patches. These overlapping patches are extracted with a stride size of 5 pixels, set empirically. The other hyperparameters have been set according to the configuration given in [35]. Fig. 5 summarises the framework of the proposed IterMiUnet, which depicts the model training on patches and the recomposition of the patches to yield the final segmented output.

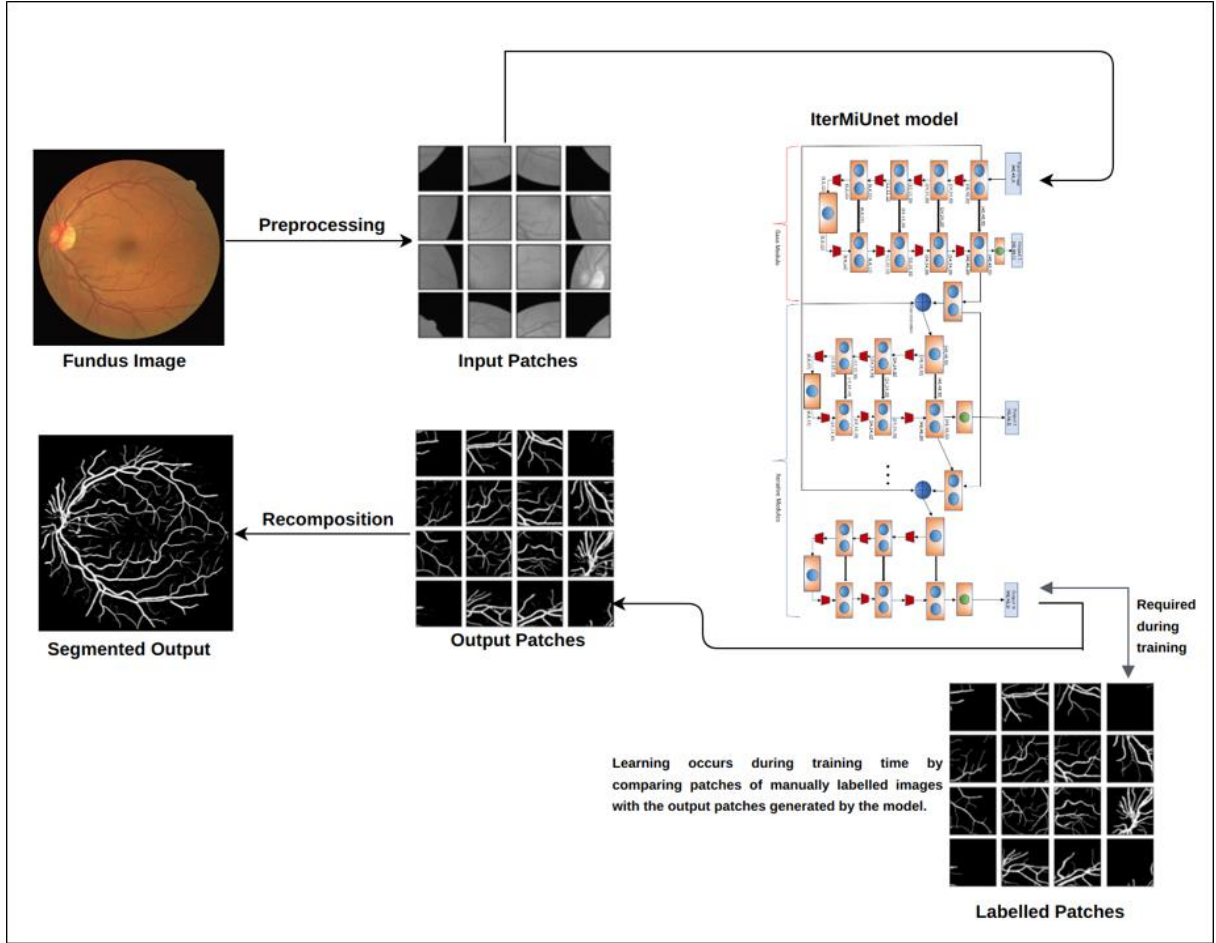

**Fig. 5** Framework diagram for the proposed IterMiUnet model

### 4.3 Evaluation Metrics

We assess the performance of the IterMiUnet model based on standard metrics, namely Accuracy (AC), Sensitivity (SE), Specificity (SP), Precision (PR), and the Area Under Curve (AUC) of the Receiver Operating Characteristic (ROC). AC measures the proportion of correctly classified pixels to the overall pixel number in the dataset. SE quantifies the fraction of correctly classified true positives. SP quantifies the fraction of correctly classified true negatives. PR measures the proportion of true positives out of all positive predicted samples in the dataset. The true-positive rate (TPR) is plotted against the false-positive rate in a receiver operating characteristic (ROC) curve. The closer the area under curve (AUC) value is to 1, the better the model's capability to distinguish between two classes. The mathematical definitions of metrics are given as:

$AC = (TP + TN)/(TP + FP + TN + FN)$ , $SE = TP/(TP + FN)$

$PR = TP/(TP + FP)$ , $SP = TN/(TN + FP)$

where TP denotes True Positives, TN denotes True Negatives, FP denotes False Positives, and FN denotes False Negatives. Furthermore, the F1 score [36], which acts as a balance of Precision and Recall (Sensitivity), is also determined. The formula is given as follows:

$F1\ score = 2TP/(2TP + FP + FN)$

### 4.4 Implementation Details

For training the model, we employed binary crossentropy as our loss function along with Adam [37] optimizer for fast convergence. Initially, the learning rate is fixed as $1e-3$. If the loss value remains the same for $m_s$ epochs, it is decreased by a factor of ten. Value of $m_s$ is set to 10. The activation function for all convolutional layers is the Rectifier Linear Unit (ReLU[38]), except for the last layer, which uses the sigmoid activation function. The dropout [39] rate is fixed as 10 percent. The model is trained for 100 epochs, and the batch size consists of 64 training samples. The value of $N$ in Iternet and IterMiUnet is fixed at 4, which means: the iterative modules are run thrice. Table 2 list the various hyperparameters values used in this study's experiments.

**Table 2** Summary of the hyperparameter configuration used in the study

| Activation function | ReLU for intermediate layers and Sigmoid for last layer |
|---|---|
| Dropout Rate | 10 % |
| Epochs | 100 |
| Batch Size | 64 |
| $N$ in Iternet & IterMiUnet | 4 |
| Patch Size | $48 \times 48$ |
| Number of Patches extracted per image | 10000 |
| Stride Size for Overlapping Patches | 5 |
| Optimizer | Adam, with an initial learning rate $1e-3$ |
| Learning rate decay factor | 10 |
| Loss Function | Binary CrossEntropy |

### 4.5 Experimental results

We compared the performance of our proposed IterMiUnet model with Unet and its variants, namely MiUnet, Recurrent Unet, and Iternet. We implemented all these models belonging to the Unet family and evaluated them across the three datasets. To compare all the Unet variants on an equal footing, all models have been trained with the same hyperparameter configuration. We used the Keras framework with TensorFlow [40] as backend on a NVIDIA Tesla k80 GPU to train our models. Secondly, we compared our proposed model with recent deep learning approaches and other segmentation approaches existing in the literature across the three datasets. Additionally, cross-training [41] and inter-rater variability [23] evaluations have also been performed for the proposed model.

### 4.5.1 Comparisons with various U-Net variants

This section discusses the comparison of our proposed IterMiUnet model against the various existing variants of Unet. By thresholding the outputs at 0.5 probability value, we obtained a binary segmentation map for each model. Tables 3 carry the evaluation results across the DRIVE, STARE, and CHASE-DB1 datasets.

The proposed IterMiUnet ranked first in terms of accuracy, F1 score, AUC, and third in sensitivity and specificity for the DRIVE dataset. Our proposed model ranked second across all metrics for the STARE dataset, except for accuracy and specificity, where we ranked third and fifth, respectively. For the CHASE-DB1 dataset, the IterMiUnet model ranked first in AUC, second in sensitivity and F1 score and third in specificity and accuracy.

The number of parameters for each model is shown in Table 4. IterMiUnet, a hybrid of Iternet and MiUnet, ranks second in terms of the lowest parameters. Although MiUnet has the lowest parameters among all the models, our proposed IterMiUnet model has consistently performed better than MiUnet in accuracy and F1 score. Even for metrics where the proposed model is ranked lower than MiUnet the decrease is very minute. IterMiUnet has just 0.15M parameters, which is considerably fewer than the Iternet's 8M parameters, but the performance is similar to the Iternet. The proposed model achieves a higher value than Iternet across two of the three datasets and a slightly lower value for sensitivity, specificity, and accuracy in the third dataset. IterMiUnet outperforms Recurrent Unet in terms of F1 score, accuracy value, and AUC across all three datasets. Although Recurrent Unet shows the highest sensitivity across all three datasets, this comes at a huge parameter cost. Recurrent Unet has the highest learnable parameters (15.7M), as shown in Table 4. In comparison, our proposed model only shows a minor decrease in sensitivity. Finally, when compared to Unet, our model has superior AUC, sensitivity, and F1 score values. In terms of specificity and accuracy, a very marginal decrease is reported.

**Table 3** Comparison of Performance of models on DRIVE, STARE, and CHASE-DB1 datasets

| DRIVE | | | | | |
|---|---|---|---|---|---|
| **Model** | **AUC** | **SE** | **SP** | **AC** | **F1 score** |
| **Unet** | 0.9788 | 0.7865 | 0.9807 | 0.9560 | 0.8198 |
| **MiUnet** | 0.9806 | 0.7972 | **0.9801** | 0.9567 | 0.8244 |
| **Recurrent Unet** | 0.9806 | **0.8097** | 0.9780 | 0.9566 | 0.8196 |
| **Iternet** | 0.9802 | 0.8087 | 0.9779 | 0.9563 | 0.8251 |
| **IterMiUnet** | **0.9810** | 0.8053 | 0.9789 | **0.9568** | **0.8262** |
| STARE | | | | | |
| **Model** | **AUC** | **SE** | **SP** | **AC** | **F1 score** |
| **Unet** | 0.9847 | 0.7892 | **0.9853** | 0.9651 | 0.8195 |
| **MiUnet** | **0.9856** | 0.8024 | 0.9840 | 0.9649 | 0.8217 |
| **Recurrent Unet** | 0.9847 | **0.8079** | 0.9833 | **0.9653** | **0.8241** |
| **Iternet** | 0.9840 | 0.7878 | 0.9849 | 0.9644 | 0.8157 |
| **IterMiUnet** | 0.9852 | 0.8069 | 0.9831 | 0.9649 | 0.8231 |
| CHASE-DB1 | | | | | |
| **Model** | **AUC** | **SE** | **SP** | **AC** | **F1 score** |
| **Unet** | 0.9812 | 0.8388 | 0.9717 | **0.9597** | **0.7891** |
| **MiUnet** | 0.9733 | 0.8442 | 0.9569 | 0.9468 | 0.7406 |
| **Recurrent Unet** | 0.9797 | **0.8556** | 0.9655 | 0.9556 | 0.7759 |
| **Iternet** | 0.9792 | 0.8188 | **0.9732** | 0.9593 | 0.7834 |
| **IterMiUnet** | **0.9812** | 0.8443 | 0.9704 | 0.9591 | 0.7875 |

**Table 4** Summarization of parameters, size, train, and inference time for each Unet variant

| | Model | | | | |
|---|---|---|---|---|---|
| | **Unet** | **MiUnet** | **Recurrent Unet** | **Iternet** | **IterMiUnet** |
| **Parameters** | 7.8 M | 0.069 M | 15.7 M | 8 M | **0.15M** |
| **Size (in MB)** | 90.2 | 1.1 | 180.5 | 94.7 | **2.3** |
| **Training time (minutes)** | 457.5 | 137.46 | 814.98 | 751.8 | **343.14** |
| **Inference time (minutes)** | 0.1625 | 0.0976 | 0.2432 | 0.1779 | **0.1259** |

Table 4 also shows the size of each model, as well as the training and inference time for each model. Each of these numbers was reported using the STARE dataset. IterMiUnet, which occupies a size of 2.3 MB, is the second smallest model after MiUnet, whereas Recurrent Unet is the largest. The fewer parameters and smaller size make IterMiUnet a lightweight model. To calculate the training time, we trained each model ten times and averaged the results. For the inference time, we reported the time it took to predict the segmentation map of a single unseen image averaged over the ten instances of each model trained previously. The averaging ensures that our stated values are a reliable estimate of training and inference time. MiUnet has the smallest parameters, model size, training, and inference time, but its oversimplification causes significant performance loss, as indicated by its performance metrics across the three datasets. The training and inference times for the remaining models are in line with expectations. IterMiUnet significantly outperforms all of the extensively parametrized models (Unet, Iternet, Recurrent Unet). IterMiunet is a better option because it preserves Iternet's excellent segmentation performance and yet remains lightly parametrized. Thus, although IterMiUnet performs similarly to existing Unet variants in terms of performance measures, as shown in Table 3. But as shown in Table 4, in terms of parameter count, size, training, and inference time, IterMiUnet performs significantly better than the heavily parametrized Unet variants.

The above discussion leads us to conclusion that IterMiUnet model not only segments the vessel structures similar to Unet and its variants, but with a significantly fewer number of learnable parameters. It saves training and inference time. The segmentation map created by various models across the three datasets can be objectively and subjectively viewed by looking at the visual segmentation results given in Fig. 6, 7, and 8. For the CHASE-DB1 dataset, MiUnet generates a vessel map with noticeable false positives. Similarly, false positives are seen in the segmentation map generated by Recurrent Unet. The thin vascular structures are missed by Unet. In comparison to others, IterMiUnet generates distinct vessel segmentation maps. Its vessel map appears to be more refined than the one produced by Iternet, with numerous disconnected vessels appearing as joint structures.

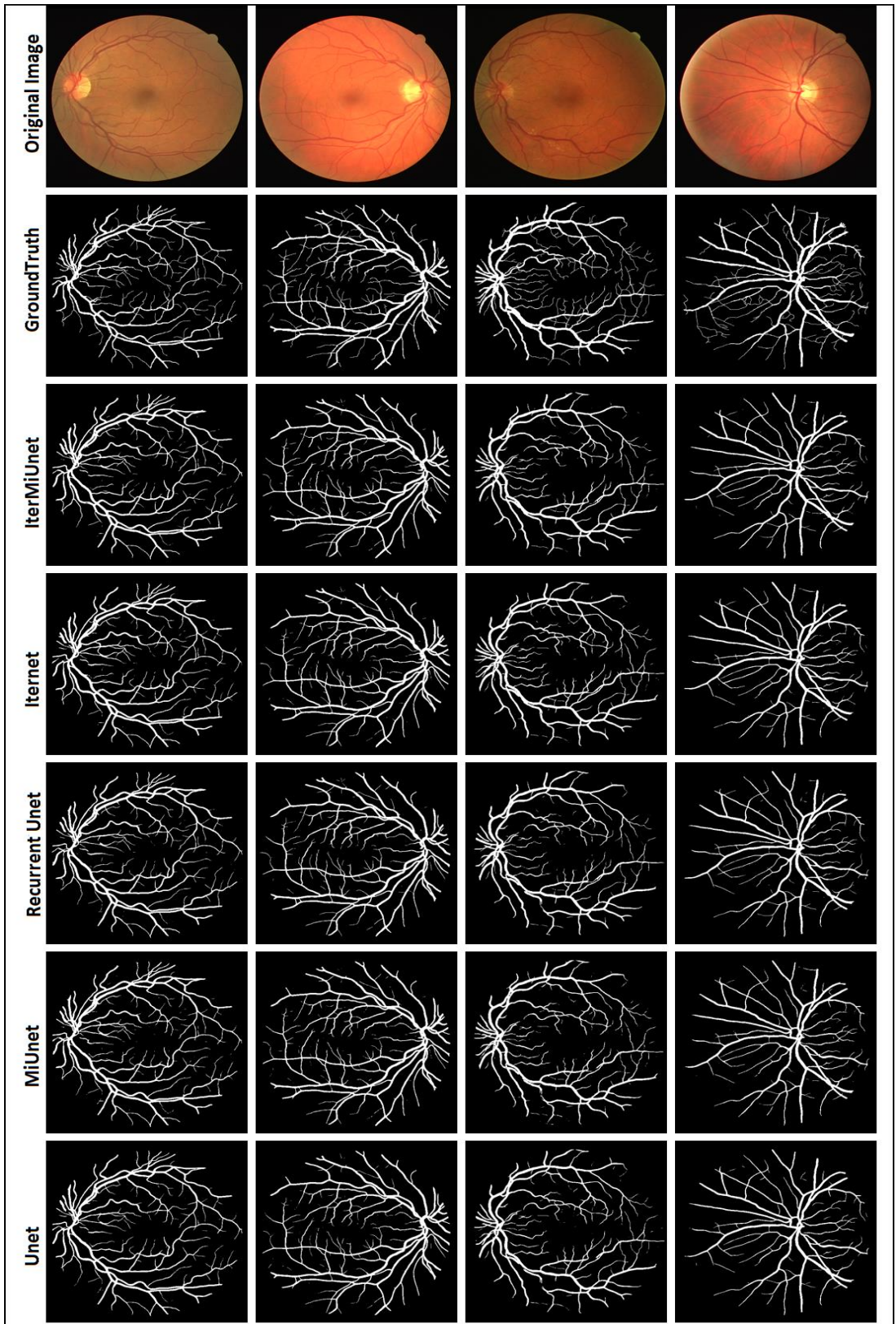

**Figure 6** Vessel Segmentation maps generated by different models on DRIVE dataset

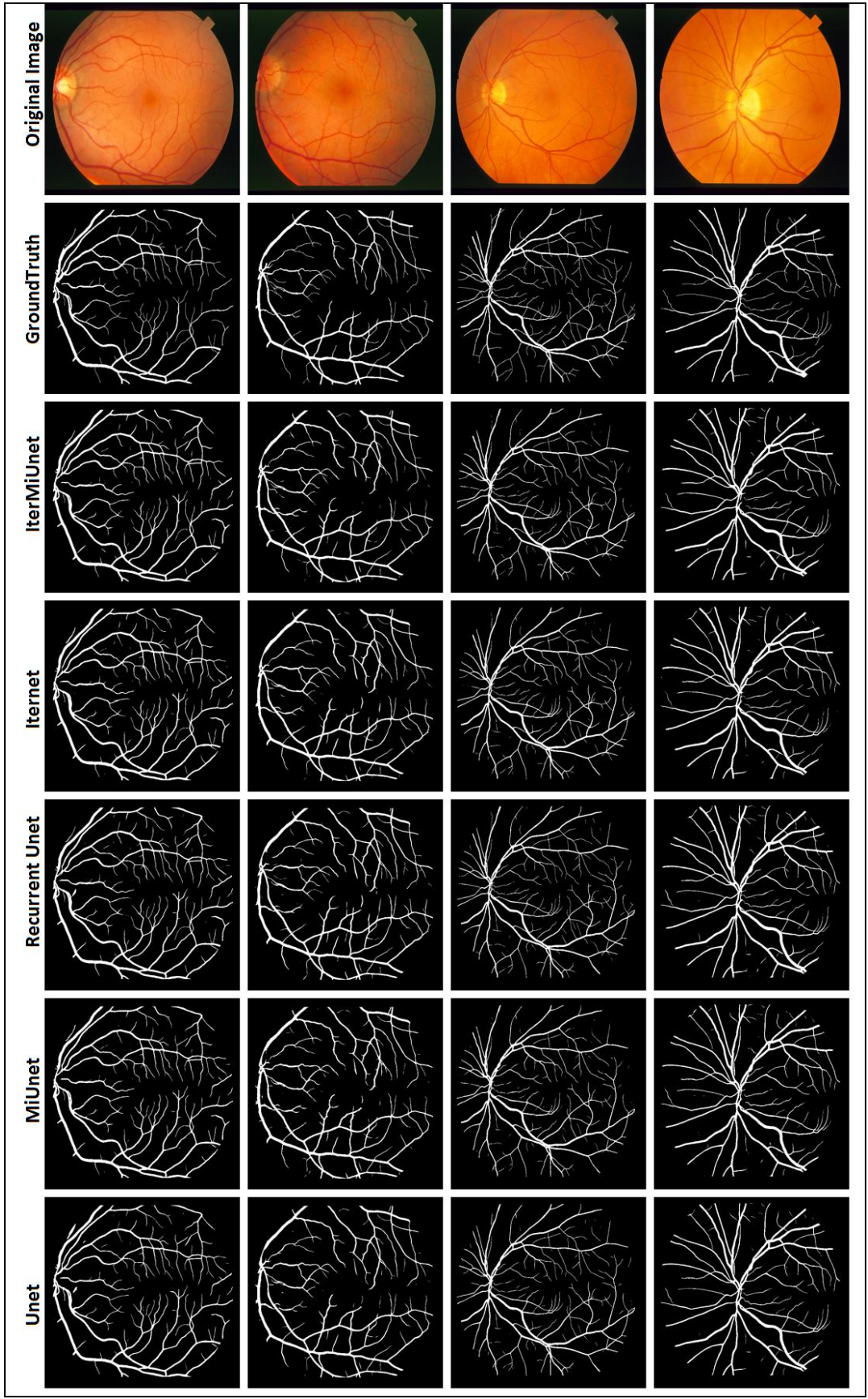

**Fig. 7** Vessel Segmentation maps generated by different models on STARE dataset

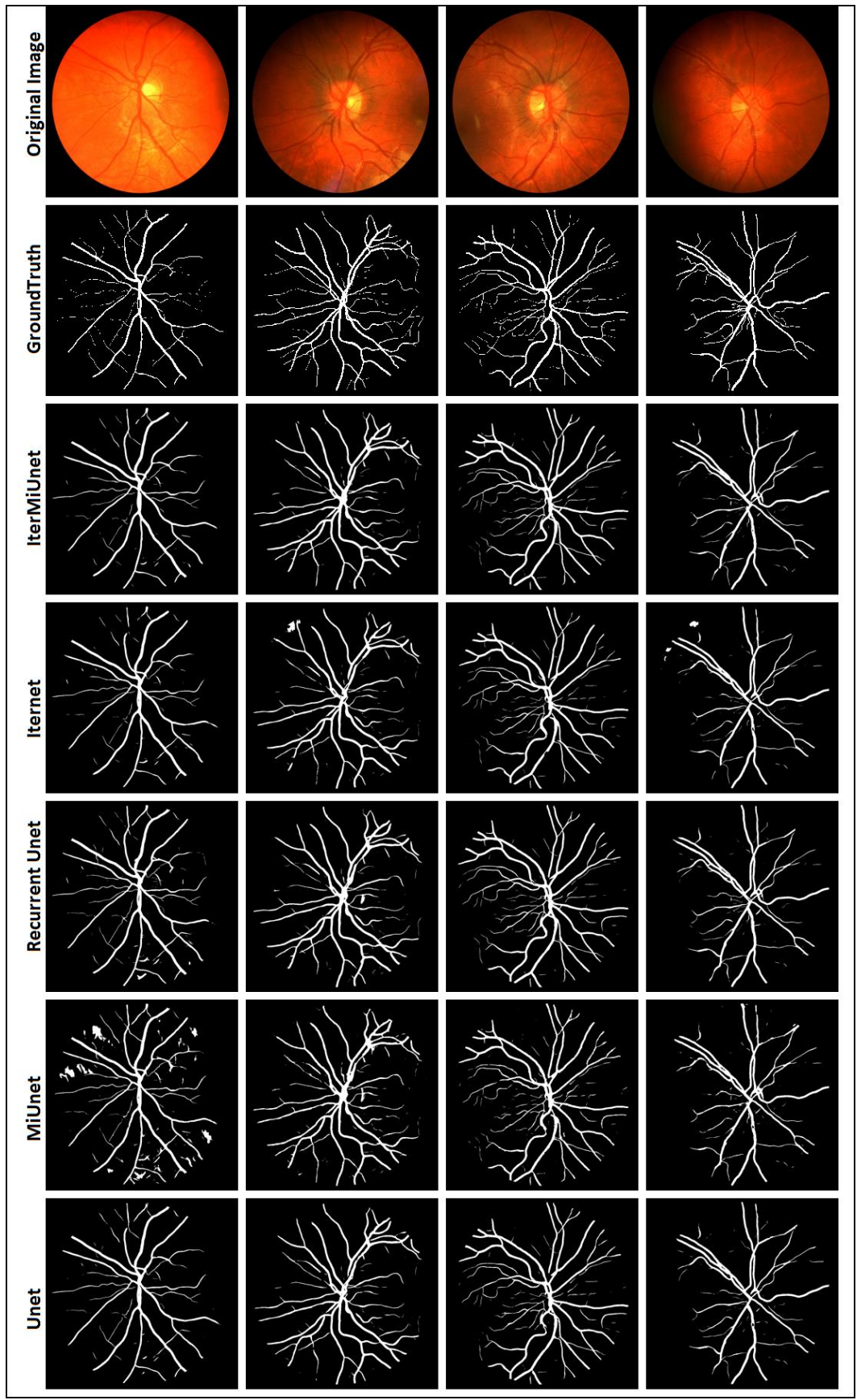

**Fig. 8** Vessel Segmentation maps generated by different models on CHASE dataset

### 4.5.2 Comparisons with existing methods

In order to further establish the efficiency of the suggested IterMiUnet model, we compare its results with other segmentation methods present in the literature, which do not necessarily use Unet or its variants to segment the retinal images. These include both unsupervised and supervised methods as well as deep learning-based methods. Table 5 depicts this comparison where all the results of previous methods were adopted from their corresponding papers. It is observed that the proposed IterMiUnet model easily outperforms the methods based on supervised and unsupervised learning in all metrics except the specificity value. In the previous section, we have already compared the proposed IterMiUnet model with other Unet-based methods and shown its excellent performance in relation to Unet-based models. Compared with deep learning-based approaches, our model achieves the second-highest sensitivity value for DRIVE and STARE, whereas the highest sensitivity value is reported for the CHASE-DB1 dataset. The proposed model attains the seventh highest specificity value for the STARE dataset, whereas it stands eighth for DRIVE and CHASE-DB1. In terms of accuracy among these various deep learning models, our model is ranked the third highest across DRIVE and STARE and fifth across CHASE-DB1. Because accuracy incorporates information about both sensitivity and specificity, it is reasonable to assert that increases in true positives for our model do not come at the expense of increased false positives. Our proposed model consistently reports an AUC value greater than 0.98 across all the datasets. Oliveira et al. [23], which reports top results across many metrics, incorporates domain information in the deep network by using wavelets and has a fairly large number of training parameters to its credit. Our proposed IterMiUnet model, on the other hand, is a very lightweight model with unprecedented fewer parameters that only uses a minimally preprocessed image as its input. Our model shows comparable performance to Oliveira et al., if not exceeding it. Among other deep learning methods, Yan et al. [24] use separate loss functions to segment thin vessels, Jin et al. [35] utilise deformable convolution to capture blood vessels at various scales, and Zhang et al. [42] incorporate complex architecture to get context and non-context features. These complex methods are easily outperformed in different metrics by the proposed IterMiUnet model. Our lightweight model reports comparable performance to the existing techniques and even exceeds them in some metrics.

**Table 5** Comparisons of the proposed model with the state-of-the-art methods

| | | DRIVE | | | | STARE | | | | CHASE-DB1 | | | |
|---|---|---|---|---|---|---|---|---|---|---|---|---|---|
| **Method** | **Year** | AUC | SE | SP | AC | AUC | SE | SP | AC | AUC | SE | SP | AC |
| **Unsupervised** | | | | | | | | | | | | | |
| Roychowdhury et al. [13] | 2015 | 0.9672 | 0.7395 | 0.9782 | 0.9494 | 0.9673 | 0.7317 | 0.9842 | 0.9560 | 0.9623 | 0.7615 | 0.9575 | 0.9467 |
| Azzopardi et al. [14] | 2015 | 0.9614 | 0.7655 | 0.9704 | 0.9442 | 0.9563 | 0.7716 | 0.9701 | 0.9497 | 0.9623 | 0.7615 | 0.9575 | 0.9467 |
| Zhang et al. [21] | 2016 | 0.9636 | 0.7743 | 0.9725 | 0.9476 | 0.9748 | 0.7791 | 0.9758 | 0.9554 | 0.9606 | 0.7626 | 0.9661 | 0.9452 |
| **Supervised** | | | | | | | | | | | | | |
| Fraz et al. [19] | 2012 | 0.9747 | 0.7406 | 0.9807 | 0.9480 | 0.9768 | 0.7548 | 0.9763 | 0.9534 | 0.9712 | 0.7224 | 0.9711 | 0.9469 |
| Roychowdhury et al. [20] | 2015 | 0.9620 | 0.7249 | **0.9830** | 0.9520 | 0.9690 | 0.7720 | 0.9730 | 0.9510 | 0.9532 | 0.7201 | 0.9824 | 0.9530 |
| **Deep Learning** | | | | | | | | | | | | | |
| Li et al. [21] | 2016 | 0.9738 | 0.7569 | 0.9816 | 0.9527 | 0.9879 | 0.7726 | 0.9844 | 0.9628 | 0.9716 | 0.7507 | 0.9581 | 0.9581 |
| Liskowski et al. [22] | 2016 | 0.9720 | 0.7763 | 0.9768 | 0.9495 | 0.9785 | 0.7868 | 0.9754 | 0.9566 | – | – | – | – |
| Oliveira et al. [23] | 2017 | 0.9821 | 0.8039 | 0.9804 | **0.9576** | 0.9905 | **0.8315** | 0.9858 | **0.9694** | 0.9855 | 0.7779 | **0.9864** | 0.9653 |
| Yan et al. [24] | 2018 | 0.9752 | 0.7653 | **0.9818** | 0.9542 | 0.9801 | 0.7581 | 0.9846 | 0.9612 | 0.9781 | 0.7633 | 0.9809 | 0.9610 |
| Jin et al. [35] | 2019 | 0.9802 | 0.7963 | 0.9800 | 0.9566 | 0.9832 | 0.7595 | **0.9878** | 0.9641 | 0.9804 | 0.8155 | 0.9752 | 0.9610 |
| Iternet [1] | 2020 | 0.9802 | **0.8087** | 0.9779 | 0.9563 | 0.9840 | 0.7878 | 0.9849 | 0.9644 | 0.9792 | 0.8188 | 0.9732 | 0.9593 |
| Uysal et al. [43] | 2021 | – | 0.7778 | 0.9784 | 0.9527 | – | 0.7558 | 0.9811 | 0.9589 | – | – | – | – |
| Zhang et al. [42] | 2022 | **0.9834** | 0.7853 | 0.9818 | 0.9565 | 0.9901 | 0.8002 | 0.9864 | 0.9668 | **0.9893** | 0.8132 | 0.9840 | **0.9667** |
| **IterMiUnet** | **2022** | **0.9810** | **0.8053** | 0.9789 | 0.9568 | 0.9852 | 0.8069 | 0.9831 | 0.9649 | 0.9812 | **0.8443** | 0.9704 | 0.9591 |

### 4.5.3 Cross Training Evaluation

It may not be practical to retrain the model on every new dataset in real-world applications. There may be instances when the model must be evaluated on a dataset other than the one for which it was trained. This is known as cross training evaluation, where the test and training datasets belong to different distributions. We have done cross training evaluations for the DRIVE and STARE datasets and compared their results with existing methods. Instead of training the models again for cross training evaluations, the IterMiUnet model used for comparison against variants of Unet as described in section 4.5.1 was used for this purpose. Table 6 summarises the cross-training results. Although the accuracy value for the DRIVE dataset was reduced from 0.9508 in cross-training evaluation, it is still the highest among all previous methods. For other metrics, namely sensitivity, specificity and AUC, we reported the second-best results. Our model gave the best results in sensitivity for the STARE dataset but performed relatively poorly across all other metrics. One possible explanation for this is that the DRIVE dataset on which the model is trained contains more fine vessels, but the test dataset of STARE has more thick vessels [41]. Further, the images in the STARE dataset are much more complicated than in the DRIVE dataset. Overall, our lightly parametrized model gives comparable performance to previous methods and even exceeds them in some metrics.

Table 6 Comparisons for Cross Training Evaluation

| Test set | Method | Year | SE | SP | AC | AUC |
|---|---|---|---|---|---|---|
| DRIVE (Trained on STARE) | Li et al. [21] | 2016 | **0.7273** | 0.9810 | 0.9486 | 0.9677 |
| | Yan et al. [24] | 2018 | 0.7014 | 0.9802 | 0.9444 | 0.9568 |
| | Jin et al. [35] | 2019 | 0.6505 | **0.9914** | 0.9481 | **0.9718** |
| | IterMiUNet | 2021 | 0.7053 | 0.9866 | **0.9508** | 0.9706 |
| STARE (Trained on DRIVE) | Li et al. [21] | 2016 | 0.7027 | 0.9828 | 0.9545 | 0.9671 |
| | Yan et al. [24] | 2018 | 0.7319 | **0.9840** | **0.9580** | **0.9678** |
| | Jin et al. [35] | 2019 | 0.7000 | 0.9759 | 0.9474 | 0.9571 |
| | IterMiUnet | 2022 | **0.7385** | 0.9703 | 0.9464 | 0.9546 |

### 4.5.4 Analysing Inter-Rater Variability

The ground truth in our vessel segmentation task is obtained by manual annotations of the fundus images done by experts. However, because this task involves subjectivity, two different observers may annotate the same fundus image differently. Especially in regions in the image where vessels are not clearly visible, annotators may mark the area as a vessel or a non-vessel, depending on their skills and previous experience. For vessels that are clearly visible, different annotators may come up with a varied estimation of their width. As a result of these factors, inter-rater variability invariably exists in the assessment process [23]. This section discusses how the segmentation results are affected by inter-rater variability. For all of the previous results presented in this work, the manual segmentation performed by the first human expert is used as the groundtruth for training and testing the model. However, for Table 7, we used the manual segmentation performed by the second expert as our groundtruth for testing the images. After training our model using the second expert's annotations, we have compared our results with the first human expert across various datasets.

Table 7 Performance comparison of the segmentation results of the IterMiUnet model when tested using the second expert's annotations as the groundtruth with the segmentation results of the first human expert's annotations

| Test Set | Method | Sensitivity | Specificity | Accuracy |
|---|---|---|---|---|
| DRIVE | First human expert | 0.8066 | 0.9674 | 0.9473 |
| | IterMiUnet | **0.8423** | **0.9799** | **0.9630** |
| STARE | First human expert | **0.6439** | 0.9883 | **0.9346** |
| | IterMiUnet | 0.6151 | **0.9915** | 0.9338 |
| CHASE-DB1 | First human expert | 0.7974 | 0.9736 | 0.9561 |
| | IterMiUnet | **0.7985** | **0.9749** | **0.9572** |

From Table 7, in the CHASE-DB1 dataset, our model outperforms the first expert in terms of sensitivity, specificity, and accuracy. Our model has a higher specificity value than the first expert in the STARE dataset, but its accuracy is marginally lower. The high specificity value in both datasets indicates the model introduced fewer false positives. It shows that there are pixels that are marked as non-vessel by our model and rejected by the first expert were confirmed to be non-vessel when the annotations of the second expert are used as ground truth for testing. The model outperformed the first expert in all the performance metrics for the DRIVE dataset. The increased value of sensitivity here shows that the model was also able to reduce false negatives in terms of the first expert. Thus, the model could even identify vessel segments marked by a second expert for the DRIVE dataset, although it was trained according to the first expert. Comparing Tables 5 and 7, we can further observe a marginal decrease in average accuracy from 0.9591 to 0.9572 for CHASE-DB1, which is understandable as the model was trained according to the annotations of the first expert. For the STARE dataset, comparing accuracy values across Tables 5 and 7 shows a significant decrease in accuracy value from 0.9649 to 0.9338, which is attributed to the considerable difference in annotations by the two experts [23]. The model seems to be closer to the second expert despite being trained according to the first expert annotations for the DRIVE dataset. This can be further corroborated by comparing the increase in accuracy from 0.9568 to 0.9630 across tables 5 and 7. Overall, there is bound to be a significant variation in the observations of the two experts, which can be attributable to their different subjective skills. In contrast, our model shows consistent performance even when evaluated against a new reference which shows its robustness to interrater variability.

### 4.6 Discussion

In a convolutional-based segmentation model, the encoder module handles the feature extraction and consists of a contracting path. The number of filters grows as the depth increases while the size of feature maps shrinks. This is because the number of low-level features (e.g., lines, tiny circles) is often quite small, but there are numerous ways to combine them into higher-level features. Thus, the end of the contracting path is characterized by *a large number of small-sized feature maps*. It is assumed that sufficient high-level abstracts regarding the context of the input image have been captured in this process. The segmentation model's decoder module consists of an expanding path. It builds on the small-sized feature maps and adds location information of various objects in the image to produce a segmented output. Unet and some of its variants like Iternet and Recurrent Unet are examples of this type of segmentation model.

There is another variant of Unet called the MiUnet that differs in the choice of filter configuration. In its encoder module, the number of filters decreases or stays the same as the depth increases. Thus, the end of the contracting path in a MiUnet-like model consists of a *small number of small-sized feature maps*. This constraint on the filters can lead to sub-optimal results because the model now has a very limited number of high-level abstracts. However, for our retinal vasculature segmentation problem, the performance of MiUnet, while far from optimal, does not deteriorate much, and it is still comparable to that of Unet. Looking at the performance metrics and the segmentation maps obtained in Table 3 and Figures 6, 7, and 8, it is clear that MiUnet still produces decent segmentation performance despite the constraints in filter configuration. This is indeed a remarkable accomplishment given that MiUnet has only 0.069M parameters, whereas Unet has about 7.8 M parameters. This indicates that the limited high-level features learned by MiUnet have successfully captured the context information, albeit the segmentation performance is still a bit inferior.

As a result, we tried to expand on the knowledge of two existing architectures, MiUnet and Iternet. We observed that the constrained filter configuration of MiUnet can still produce manageable results and investigated ways to improve its segmentation performance. Our emphasis on the usage of MiUnet-like architecture also stems from the fact that almost all existing state-of-the-art models possess huge parameters, making them complex and vulnerable to overfitting. Iternet, on the other hand, has an iterative module that can refine the imprecise segmentation map in an end-to-end manner and performs significantly better than Unet on our segmentation task. Therefore, we incorporated the lightweight MiUnet structure inside the Iternet architecture, producing a new model called IterMiUnet. Despite leveraging the MiUnet network within it, the new model does not exhibit poor segmentation performance. This is due to the usage of iterative modules, which ensures that the segmentation map is appropriately improved. Furthermore, using this constrained filter configuration results in a model that is also lightly parametrized. This distinguishes IterMiUnet from Iternet, even though both have the same complex structure consisting of a base and iterative module. Therefore, IterMiUnet can overcome the limitations of both Iternet and MiUnet, resulting in an effective segmentation model with much fewer parameters.

### 5. Conclusion

Unet architecture has shown to be an excellent choice for semantic segmentation and retinal vasculature segmentation. However, the usefulness of Unet architecture is limited by the network's depth, and it cannot appropriately segment the intricate vasculature of retinal images. Therefore, several variants of Unet have been proposed from time to time to overcome its limitations. The major problem with Unet and its variants is that they are heavily parametrized. As a result, Unet-based models have fallen out of favor. This study proposes a new model for retinal vascular segmentation termed IterMiUnet. The model is based on the Unet structure but has just about 0.15 M parameters, making it a lightly parametrized model. IterMiUnet could be very useful in the medical domain, a field that invariably suffers from data scarcity due to which heavily parametrized models tend to overfit. The model does not compromise the network's depth to reduce the parameters. The new model has been tested across three mainstream datasets, i.e., DRIVE, STARE, and CHASE-DB1. Although being very lightly parametrized, the model's performance is comparable to the existing segmentation approaches to retinal images. This shows the generalization capability of our model. The cross-training evaluations conducted show the robustness of the proposed model for the training set. Our model also shows robustness to inter-rater variability. The proposed model has a significantly low training and inference time, making it suitable for application in diagnostic systems.

One of the shortcomings of our proposed model is the inaccurate segmentation of very thin vessels. Further attempts will be made in future work to incorporate domain knowledge of the vessels, including their geometrical structure, into our network. This will help our model get around the complex blood vessel structures and further improve the vessel map by emphasizing thin vessels, which are harder to extract, even with the proposed IterMiUnet model.